\documentclass[sigconf,natbib,nonacm]{acmart}
\usepackage{fdiaz}
\usepackage{algorithm}
\usepackage{algpseudocode}

\usepackage{booktabs}

\setcopyright{rightsretained}

\usepackage{subcaption}

\acmDOI{10.475/123_4}

\acmISBN{123-4567-24-567/08/06}

\acmConference[WOODSTOCK'97]{ACM Woodstock conference}{July 1997}{El
  Paso, Texas USA}
\acmYear{1997}
\copyrightyear{2016}

\acmArticle{4}
\acmPrice{15.00}

\newcommand{\query}[0]{q}
\newcommand{\queries}[0]{\mathcal{Q}}

\newcommand{\numdocs}[0]{n}

\newcommand{\numrel}[0]{m}

\newcommand{\rankings}[0]{S_{\numdocs}}

\newcommand{\metric}[0]{\mu}

\newcommand{\lexiprecision}[0]{\text{LP}}
\newcommand{\sgnlexiprecision}[0]{\text{sgnLP}}
\newcommand{\rrlexiprecision}[0]{\text{rrLP}}

\newcommand{\lexiprecisionpref}[0]{\succ_{\lexiprecision}}
\newcommand{\rrpref}[0]{\succ_{\rr}}

\newcommand{\rr}[0]{\mathrm{RR}}
\newcommand{\drr}[0]{\delta\rr}

\newcommand{\expectation}[2]{\mathbb{E}_{#1}\left[#2\right]}

\newcommand{\esl}[0]{\text{ESL}}

\newcommand{\rlone}[0]{\text{RL}_1}

\newcommand{\ndocs}[0]{n}

\newcommand{\rlx}[0]{\pi}
\newcommand{\rly}[0]{\pi'}

\newcommand{\RP}[0]{p}
\newcommand{\RPx}[0]{\RP}
\newcommand{\RPy}[0]{\RP'}

\newcommand{\sgn}[0]{\text{sgn}}

\newcommand{\preferencefn}[0]{\Delta}
\newcommand{\latentpreferencefn}[0]{\hat{\preferencefn}}

\definecolor{BrickRed}{HTML}{B6321C}
\newcommand{\better}[1]{\textcolor{BrickRed}{#1}}
\newcommand{\best}[1]{\textcolor{BrickRed}{\textbf{#1}}}

\begin{document}
\title{Best-Case Retrieval Evaluation: Improving the Sensitivity of Reciprocal Rank with Lexicographic Precision}
\author{Fernando Diaz}
\orcid{0000-0003-2345-1288}
\affiliation{%
  \institution{Google}
  \city{Montr\'eal}
  \state{QC}
  \country{Canada}
}
\email{diazf@acm.org}

\begin{abstract}
    Across a variety of ranking tasks, researchers use reciprocal rank to measure the effectiveness for users interested in exactly one relevant item.  Despite its widespread use, evidence suggests that reciprocal rank is brittle when discriminating between  systems.  This brittleness, in turn, is compounded in modern evaluation settings where current, high-precision systems may be difficult to distinguish.  We address the lack of sensitivity of reciprocal rank by introducing and connecting it to the concept of best-case retrieval, an evaluation method focusing on assessing the quality of a ranking for the most satisfied possible user across possible recall requirements.  This perspective allows us to  generalize reciprocal rank and define a new preference-based evaluation we call lexicographic precision or lexiprecision.  By mathematical construction, we ensure that lexiprecision preserves differences detected by reciprocal rank, while empirically improving sensitivity and robustness across a broad set of retrieval and recommendation tasks.  
\end{abstract}

\maketitle
\section{Introduction}
\label{sec:introduction}
Evaluating ranking systems for users seeking exactly one relevant item has a long history in information retrieval.  As early as 1968, \citet{cooper:esl} proposed \textit{Type 1 expected search length} or $\esl_1$, defined as the rank position of the highest ranked relevant item.  In the context of TREC-5, \citet{kantor:rr} proposed using the reciprocal of $\esl_1$ in order to emphasize rank changes at the top of the ranked list and modeling the impatience of a searcher as they need to scan for a single item.  Over the years,  reciprocal rank (and less so $\esl_1$) has established itself as a core metric for retrieval \cite{diaz:neurips-2020-tutorial} and recommendation \cite{castells:offline-recsys-eval}, adopted in situations where there is actually only one relevant item as well as in situations where there are multiple relevant items.  Given two rankings,  reciprocal rank and $\esl_1$ always agree in terms of which ranking is better.  Because of this, we refer to them collectively as the recall level 1 or $\rlone$ metrics.

Despite the widespread use of reciprocal rank, recent evidence suggests that it may brittle when it comes to discriminating between ranking systems \cite{valcarce:recsys-ranking-metrics-conf,valcarce:recsys-ranking-metrics-journal,ferrante:interval-scale-metrics}.  In particular, the low number of unique values of reciprocal rank means that, especially when evaluating multiple highly-performing systems, we are likely to observe tied performance.   \citet{voorhees:too-many-relevants} demonstrate that these conditions exist in many modern deep learning benchmarks.

We address these issues  by theoretically interpreting $\rlone$ as  a population-level metric we refer to as \textit{best-case retrieval evaluation}.  This allows us to propose a generalization of the $\rlone$ ordering based on social choice theory \cite{sen:collective-choice-and-social-welfare} and preference-based evaluation \cite{diaz:rpp}.  This evaluation method, lexicographic precision or lexiprecision, preserves any strict ordering between rankings based on $\rlone$ while also providing a theoretically-justified ordering when $\rlone$ is tied. 

\begin{figure}
    \includegraphics[width=0.60\linewidth]{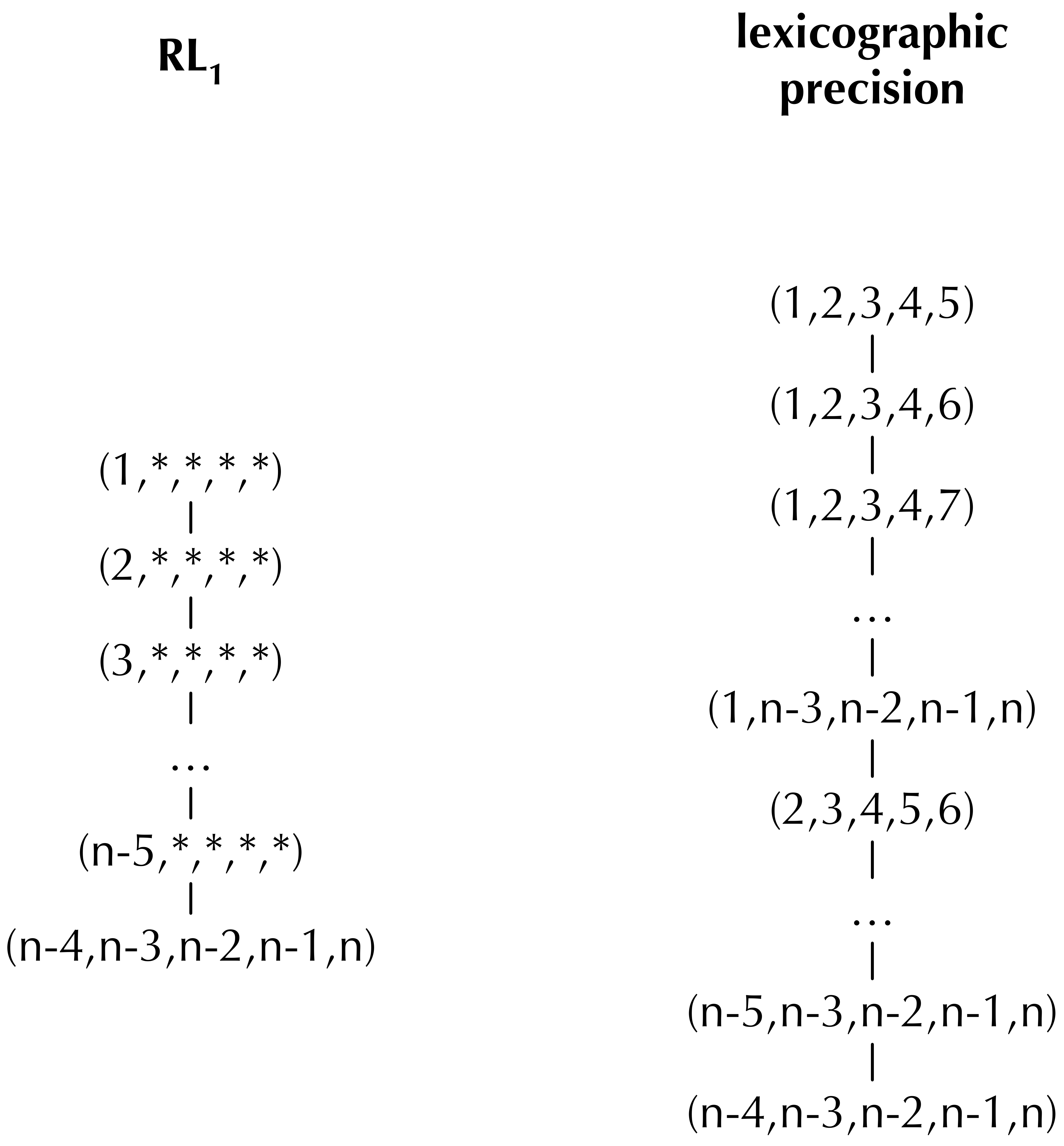}
    \caption{Hasse diagram of possible positions of relevant items.  Each tuple represents the possible positions of five relevant items in a corpus of $\numdocs$ items.  $\rlone$ metrics such as reciprocal rank (left) have $\numdocs-4$ unique values and, therefore, result in a \textit{partial order} over all possible positions of  relevant items. Lexicographic precision (right) is a \textit{total order} over all possible positions of  relevant items that preserves all strict orders in $\rlone$ evaluation. }\label{fig:hasse}
\end{figure}

We compare lexiprecision and $\rlone$ orderings using  Hasse diagrams in Figure \ref{fig:hasse}.  On the left, we show the partial order of all possible positions of five relevant items in a corpus of size $\numdocs$.  Since reciprocal rank and $\esl_1$ only consider the position of the first relevant item,  we only have $\ndocs$ different relevance levels.  While this may not be an issue in general (since $\ndocs$ is usually large), the number of rankings within each level can be very large and multiple highly effective systems can result in numerous ties.  In contrast, lexiprecision has one relevance level for each unique arrangement of relevant items.  That is, the number of relevance levels scales with the number relevant items and, by design, two rankings are tied \textit{only} if they place relevant items in exactly the same positions.

In this paper,  we contribute to the theoretical understanding of evaluation through a detailed study of $\rlone$ metrics, best-case retrieval evaluation, and lexiprecision.  In Section \ref{sec:motivation}, we motivate our work by showing that $\rlone$ has fundamental theoretical limits, especially in situations where there are multiple relevant items.  In Section \ref{sec:algorithms}, we   demonstrate that $\rlone$ can be interpreted as best-case retrieval evaluation, allowing us to to address its limitations by using methods from social choice theory and generalizing it as lexiprecision.  In Section \ref{sec:results}, we then conduct extensive empirical analysis to show that lexiprecision is strongly correlated with $\rlone$ metrics while substantially improving its discriminative power.\footnote{In lieu of an isolated `Related Work' section, we have included discussion of relevant literature when necessary.  This helps make connections explicit to our work.   }

\section{Motivation}
\label{sec:motivation}
Our work is based on the observation that  ceiling effects are inherent in $\rlone$ evaluation.  Assume a standard ranking problem where, given a query with $\numrel$ associated relevant items, a system orders all $\numdocs$ documents in the collection in decreasing order of predicted relevance.  
The set of all possible rankings of $\numdocs$ is referred to as the symmetric group over $\numdocs$ elements and is represented as $\rankings$.  
For a given ranking $\rlx\in\rankings$, let $\RPx_i$ be the position of the $i$th highest-ranked relevant item.  We can then define reciprocal rank as $\rr_1(\rlx)=\frac{1}{\RPx_1}$.  When no relevant document is retrieved  (e.g. if no relevant items are in the system's top $k$ retrieval), we set $\rr_1=0$.  For two rankings, we define $\drr_1(\rlx,\rly)=\rr_1(\rlx)-\rr_1(\rly)$.  For the remainder of this section, we will use reciprocal rank for clarity although the analysis applies to $\esl_1$ as well.  

Although we can easily see that there are $\numdocs$ different values for $\rr_1(\rlx)$, we are interested in the  distribution of ties amongst system rankings for these $\numdocs$ values as predicted by theoretical properties of reciprocal rank.  Specifically, we want to compute, for a given position of the first relevant item $\RPx_1$ and a \textit{random} second ranking,  the probability that we will observe a tie.  For any $\RPx_1$, there are $\numdocs-\RPx_1\choose\numrel-1$ tied arrangements of positions of relevant items amongst all of the possible arrangements $\RPy$ from a second system.  If we sample an arrangement of relevant items uniformly at random, then the probability of a tie with $\rlx$ is $Pr(\RPx_1=\RPy_1|\RPx_1)=\frac{{\numdocs-\RPx_1\choose\numrel-1}}{{\numdocs\choose\numrel}}$.

\begin{figure}
    \includegraphics[width=0.80\linewidth]{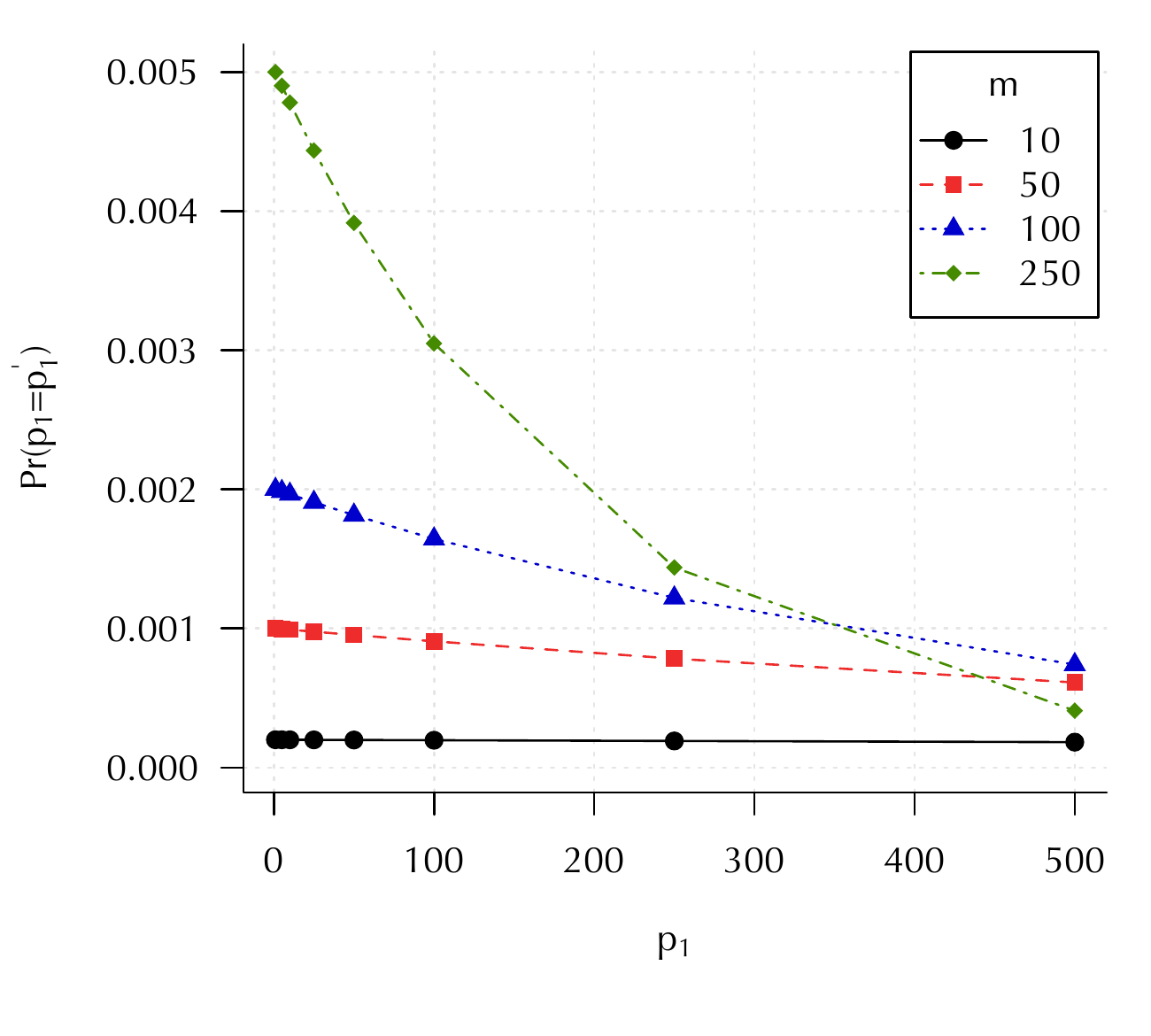}
    \caption{Given a ranking where the highest-ranked relevant item is at position $\RPx_1$, the probability of a tie with a second ranking sampled uniformly from all arrangements of relevant items for a corpus size of $\numdocs=50000$.  This figure and others are best rendered or printed in color.}\label{fig:motivation:ptie}
\end{figure}

 We plot this probability in Figure \ref{fig:motivation:ptie}.  We can observe that, when we have few relevant items (i.e. small $\numrel$), we have a relatively small and uniform probability of ties across all values of $\RPx_1$.  However, as we increase the number of relevant items, the distribution begins to skew toward a higher probability of a tie as $\RPx_1$ is smaller.  This means that, if we have a ranking where the first relevant item is close to the top, even if the second ranking is drawn uniformly at random, we will be more likely to find a tie than if the first relevant item were lower in the ranking.

\begin{figure}
    \includegraphics[width=0.80\linewidth]{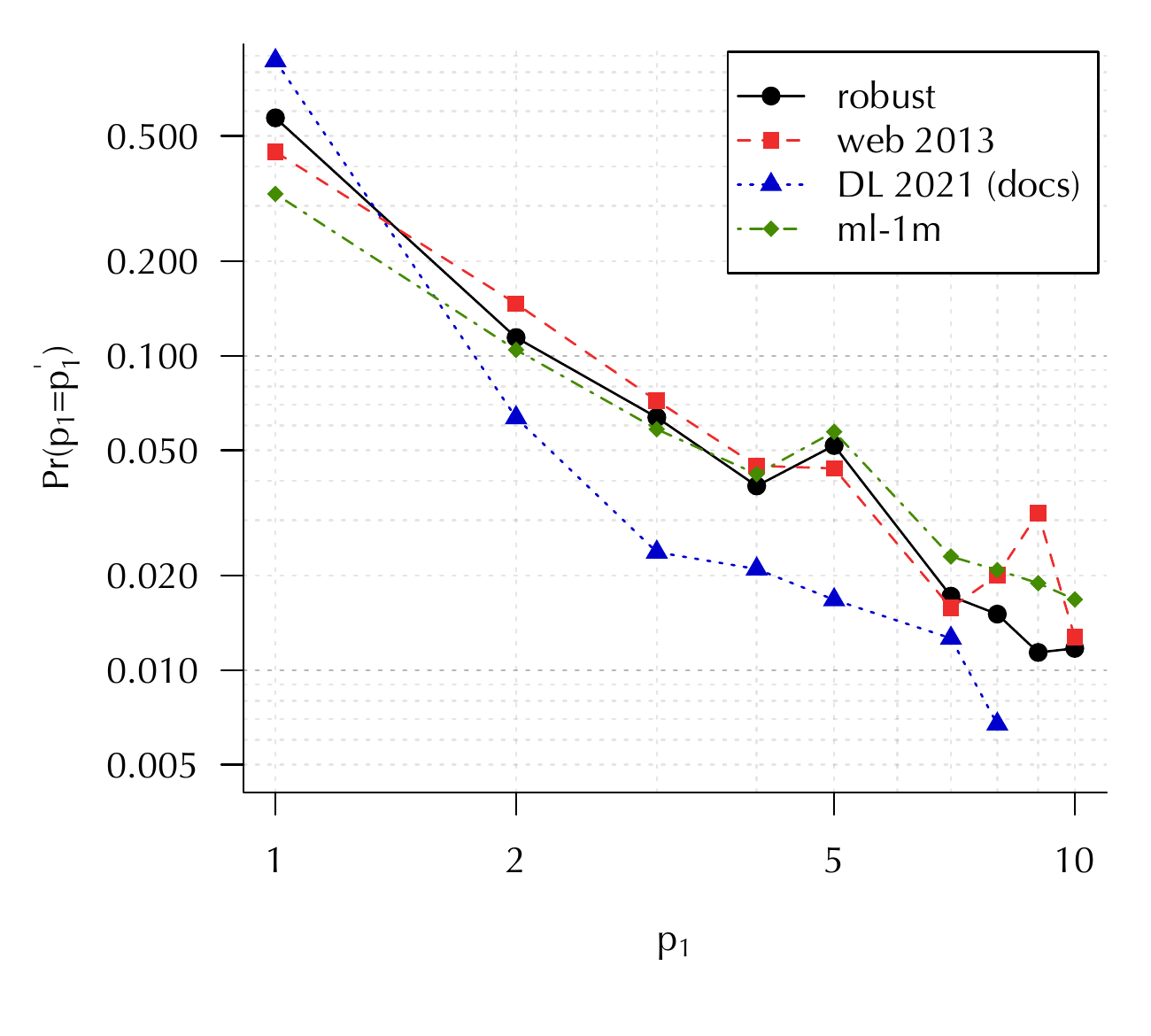}
    \caption{Empirical probability of a tie with  a second ranking for several benchmarks (see Section \ref{sec:data} for details).  Horizontal and vertical axes are on a logarithmic scale for clarity. }\label{fig:motivation:ceiling-effect}
\end{figure}

While our analysis indicates a lack of sensitivity of reciprocal rank for $\RPy$ drawn uniformly at random as $\numrel$ increases, we are also interested in the probability of  ties when $\RPy$ is drawn from rankings produced by real systems.  We collected runs associated with multiple public benchmarks (see Section \ref{sec:data} for details) and computed the the empirical distribution of ties conditioned $\RPx_1$ (Figure \ref{fig:motivation:ceiling-effect}). Because of the highly skewed distribution, we plot the logarithmic transform of the probability of a rank position.  As we can see, across both older and newer benchmarks, the probability of a tie for rankings when the top-ranked relevant item is at position 1 is substantially larger than if we assume $\RPy$ is drawn uniformly at random.  The 2021 TREC Deep Learning track data in particular demonstrates higher skew than others, confirming observations previously made about saturation at top rank positions \cite{voorhees:too-many-relevants}.

Taken together, these results demonstrate a fundamental limitation of $\rlone$ metrics (i.e., reciprocal rank and $\esl_1$) for evaluation.  As retrieval and other scenarios where reciprocal rank is used begin to attract highly performant systems, we need to extend our evaluation approaches to address these issues.

\section{Lexicographic Precision}
\label{sec:algorithms}
$\rlone$ evaluation emphasizes precision by considering the position of the top-ranked relevant item and ignoring the positions of other relevant items.  However, \textit{only} ever looking at the position of the top-ranked relevant item results  in the ceiling effects described in the previous section.  Our goal is to develop an evaluation method that preserves the ordering of a pair of rankings by $\rlone$ (i.e., agree with $\rr_1$ when $\rr_1(\rlx)\neq\rr_1(\rly)$) and provides a justified ordering of a pair of rankings when $\rlone$ is tied (i.e., generate a sensible order when $\rr_1(\rlx)=\rr_1(\rly)$). Although metrics like expected reciprocal rank \cite{chapelle:err} and average precision include reciprocal rank as a component in their computation, they are not \textit{guaranteed} to preserve the ordering of reciprocal rank when there is one.  In this section, we will interpret $\rlone$ metrics as best-case retrieval evaluation, allowing us to derive a preference-based evaluation method based on social choice theory.  

\subsection{Best-Case Retrieval Evaluation}
\label{sec:algorithms:best-case}
When a user approaches a retrieval system, there is a great deal of uncertainty about their information need.  
While a request such as a text query provides information about  which items in the corpus might be relevant, it says much less about the user's appetite for relevant information.  As a result, there is an implicit population of possible users issuing any particular request, each of whom may have a different utility for any particular ranking.  In this section, we explore two types of uncertainty and demonstrate that, from both perspectives, $\rlone$ evaluation represents the best-case utility over that population.

We first consider uncertainty over recall requirements. \citet{robertson:ap-user-model} presented a model for evaluating rankings based on the diverse set of recall requirements that a user might have.  Given a request and its associated relevant items,  users may be interested in one relevant item, a few relevant items, or the complete set of all relevant items.  We can assess the quality of a ranking for any particular user recall requirement with what \citet{cooper:esl} refers to as the \textit{Type 2 expected search length}: the number of items a user with requirement $i$ has to scan before finding $i$ relevant items.  So, each information need has $\numrel$ recall levels and $\rlone$ is the evaluation measure associated with users  requiring exactly one relevant item.  From this perspective, we can, for a specific ranking, look at how utility is distributed amongst possible users, as represented by their recall levels.  For example, we can ask how utility for users with high and low recall requirements compares; or what the average utility across these populations is.  While previous work has looked at the average-case utility \cite{diaz:rpp} and worst-case utility \cite{diaz:lexirecall:arxiv}, in this work we suggest that $\rlone$ represents the best-case performance over these possible users.  The proof is relatively simple.  Because $\rr_i$ monotonically degrades in rank,  the best-case utility over this representation of users is $\rr_1$ (equivalently $\esl_1$).  The next-best-case is $\rr_2$ and so forth until we reach $\rr_\numrel$, which we refer to as the worst-case. So, given two rankings $\rlx$ and $\rly$, observing $\rr_1(\rlx)>\rr_1(\rly)$ implies that the best-case performance over possible user recall requirements is higher in $\rlx$ compared to $\rly$.

Next, we consider uncertainty over psychologically relevant items.  When evaluating a retrieval system, we often use relevance labels derived from human assessors or statistical models.  But what if a specific user does not find the top-ranked  item labeled relevant actually relevant to them?  For example, a user may have already seen a specific item or they may desire an item with a specific (missing) attribute.  A judged relevant item might be inappropriate for any number of reasons not expressed in the request.  The concept of \textit{psychological relevance} \cite{harter:psychological-relevance} suggests that judging any item relevant \textit{in general} (as is the case in many retrieval benchmarks, including those used in TREC) is a necessary but not sufficient criteria to determine an item's psychological relevance to any particular user.  From this perspective, there are $2^\numrel-1$ possible non-empty sets of relevant items for a specific request, each representing psychological relevance to a possible user.  Nevertheless, amongst these possible users,  if they are interested in precisely one relevant item, there are $\numrel$ unique utilities.  Again, since $\rlone$ monotonically decreases in rank, the best-case utility is $\rr_1$, followed by $\rr_2$ until we reach $\rr_\numrel$.

Both uncertainty over recall levels and over psychological relevance focus on possible populations of users.  Because the utility to the user implies utility to the system designer (e.g., for objectives like retention), understanding the best-case performance is valuable in decision-making.  From the perspective of social choice theory, best-case retrieval evaluation is inherently optimistic and represents risk-seeking decision-making.

\subsection{Lexicographic Precision}
\label{sec:algorithms:lexiprecision}
The problem with evaluating for best-case retrieval (as shown in Section \ref{sec:motivation}) is the tendency for multiple rankings to be tied, especially as 
\begin{inlinelist}
    \item we increase the number of relevant items and 
    \item systems optimize for retrieval metrics.  
\end{inlinelist}  We can address these ceiling effects by developing a best-case  \textit{preference-based evaluation} that focuses on measuring differences in performance instead of absolute performance \cite{diaz:rpp}.  While \textit{metric-based evaluation} models the preference between rankings by first computing some evaluation metric for each ranking, preference-based evaluation explicitly models the preference between two rankings.  Prior research has demonstrated that preference-based evaluation can be much more sensitive than metric-based evaluation \cite{diaz:rpp}, making it well-suited for addressing the ceiling effects described in Section \ref{sec:motivation}.

Under best-case preference-based retrieval, we are interested in answering the question, `under the best possible scenario, which ranking would the user prefer?'  In this respect, it is a user-based evaluation method, but one based on preferences and measurement over a population of users.  More formally, given an information need and two rankings $\rlx$ and $\rly$ associated with two systems, metric-based evaluation uses an evaluation metric $\metric : \rankings\rightarrow\Re$ (e.g. reciprocal rank or average precision) to compute a preference,
\begin{align*}
    \metric(\rlx) > \metric(\rly) &\implies \rlx \succ \rly     
\end{align*}
where $\rlx \succ \rly$ indicates that we prefer $\rlx$ to $\rly$.  Notice  that, if $\metric(\rlx) = \metric(\rly)$, then we cannot infer a preference between $\rlx$ and $\rly$.  
We contrast this with preference-based evaluation, which directly models this relationship $\preferencefn: \rankings\times\rankings\rightarrow\Re$,
\begin{align*}
    \preferencefn(\rlx,\rly) > 0 &\implies \rlx \succ \rly 
\end{align*}

\begin{figure}
    \centering
    \begin{subfigure}[b]{0.45\columnwidth}
        \centering
        \includegraphics[width=0.80\linewidth]{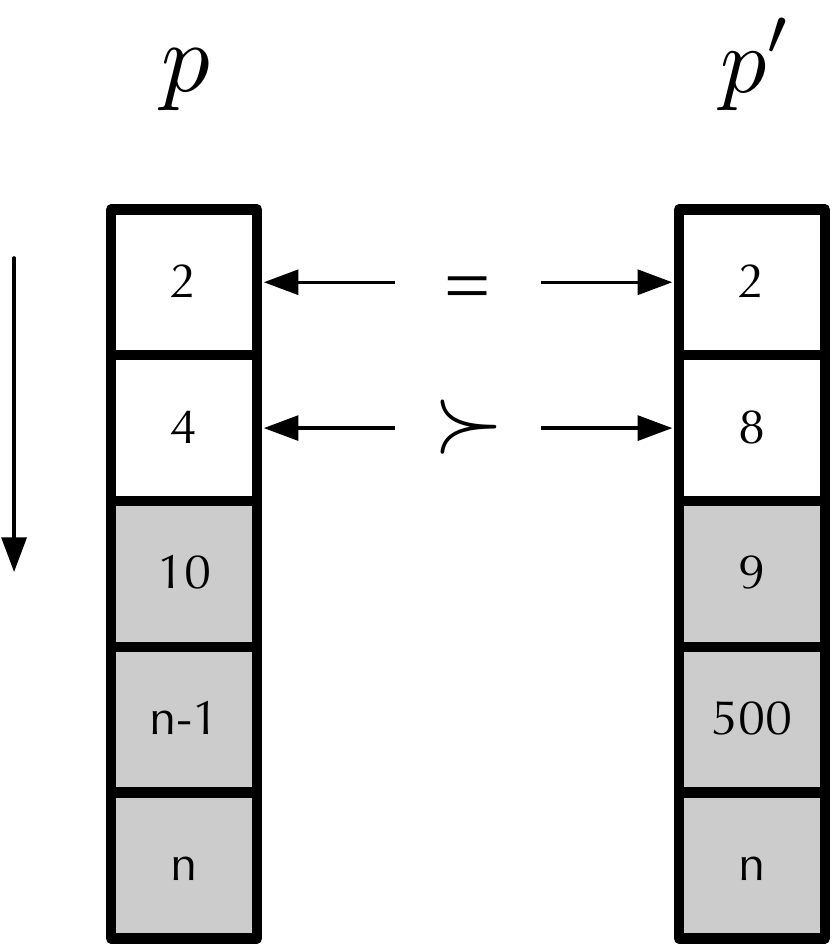}
        \caption{Lexicographic Precision}\label{fig:lexiprecision:lexiprecision}
    \end{subfigure}\hspace{.25in}\begin{subtable}[b]{0.45\columnwidth}
        \centering
        \begin{tabular}{lc}
            &$\Delta(\pi,\pi')$\\
            \hline
            $\delta\text{RR}_1(\pi,\pi')$&$\phantom{-}0$\\
            $\delta\text{ESL}_1(\pi,\pi')$&$\phantom{-}0$\\
            $\text{sgnLP}(\pi,\pi')$&$-1$\\
            $\text{rrLP}(\pi,\pi')$&$-\frac{1}{8}$\\
            \\
        \end{tabular}
        \caption{Preference Magnitude }\label{fig:lexiprecision:magnitude}
    \end{subtable}
    \caption{Lexicographic precision between two rankings $\rlx$ and $\rly$ with $\numrel=5$ relevant items in corpus of size $\numdocs$.  (\subref{fig:lexiprecision:lexiprecision}) Using the sorted positions of relevant items, lexiprecision returns a preference based on the  highest-ranked difference in positions.  (\subref{fig:lexiprecision:magnitude}) The magnitude of preference between $\rlx$ and $\rly$ under different schemes.  }\label{fig:lexiprecision}
\end{figure}

Our goal is to design a preference-based evaluation that preserves the best-case properties of $\rlone$ metrics with much higher sensitivity.  Consider the two position vectors $\RPx$ and $\RPy$ in Figure \ref{fig:lexiprecision} associated with the two rankings $\rlx$ and $\rly$.

These two vectors are tied in the best case (i.e., $\RPx_1=\RPy_1$). 
However, we can break this tie by looking at the next-best case (i.e. $\RPx_2$) where, because $\RPx_2<\RPy_2$, we say that $\rlx\succ\rly$.  If we had observed a tie between the next-best case, we could compare $\RPx_3$, and so forth.  This is known as lexicographic sorting in the social choice literature \cite{sen:collective-choice-and-social-welfare} and reflects a generalization of best-case sorting.  Given two sorted vectors of utilities, here reflected by the rank position, the lexicographic maximum begins by looking at utilities in the best-off positions (i.e. $\RPx_1$ and $\RPy_1$) and iteratively inspects lower utility positions until we find an inequality.  

If  we exhaust all $\numrel$ relevance levels, we indicate that there is not preference between the rankings.  Note that a tie can only happen if two rankings have all relevant items in exactly the same positions.   Lexicographic sorting generates a total ordering over all positions of relevant items, in contrast with just inspecting $\RPx_1$, which compresses all arrangements onto $\numdocs$ possible values.  Because of its basis in lexicographic ordering, we refer to this lexicographic precision or lexiprecision.

\subsection{Number of Ties Under Lexicographic Precision}
\label{sec:algorithms:numties}
We can contrast the number of ties as $\numrel$ increases in $\rlone$ metrics with the number of ties as $\numrel$ increases in lexiprecision.  In the latter, we only observe ties when the positions of the relevant items for two rankings are the same and, therefore,  we have $\numdocs\choose\numrel$ possible `values' and the number of ties given a fixed ranking is constant.  If we add $k$ relevant items, the number of `values' \textit{increases}, resulting in an increase in discriminative power.  Specifically, if we add $k$ relevant items to $\numrel$, then the number of possible values scales exponentially in $k$. 
\begin{align}
    \frac{{\numdocs\choose\numrel+k}}{{\numdocs\choose\numrel}}&=
    \prod_{i=\numrel+1}^{\numrel+k} \frac{\numdocs+1-i}{i}
\end{align}
By contrast, for $\rlone$ metrics, this increase in the number of unique position vectors needs to be allocated to a fixed $\numdocs$ values, resulting in collisions, as suggested by the pigeonhole principle.  Moreover, these collisions will tend to increasingly occur at values associated with position vectors where $\RPx_1$ is small (Section \ref{sec:motivation}).  

\subsection{Best-Case Retrieval Evaluation Revisited}
\label{sec:algorithms:best-case-lexiprecision}
In Section \ref{sec:algorithms:best-case}, we described two dimensions of uncertainty in retrieval evaluation: recall level and psychological relevance.  In both cases, we saw that the best-case utility was represented by $\rlone$.  In terms of preference-based evaluation, we would like to show that, for both recall level uncertainty and psychological relevance uncertainty, the highest ranked difference in utility will be $\drr_{i^*}$, where $i^*=\argmin_{j\in[1,\numrel]}\drr_i(\rlx,\rly)\neq0$. This is clear for recall level uncertainty  because the population of possible users exactly matches the recall levels defining $i^*$.  

However, for psychological relevance uncertainty, we have $2^\numrel-1$ possible users.  That said, there are only $\numrel$ possible $\rlone$ metric values.  Moreover, the number of possible users tied at the first recall level is $2^{\numrel-1}$; at the second recall level is $2^{\numrel-2}$; down to the final recall level where there is a single possible user.  This arrangement of ties is the same regardless of the exact positions of the relevant items.  Therefore, if we observe $\drr_1=0$, we will observe $2^{\numrel-1}$ ties amongst the possible psychological relevance states where where the first relevant item is at position $\RPx_1$.  The next highest utility is, by the monotonicity of $\rlone$ metrics, associated with the second recall level.  We can continue this procedure until we observe an inequality, which will occur exactly at the first $i$ such that $\drr_i(\rlx,\rly)\neq0$.  In other words, $i^*$.  

These observations are important since they demonstrate that lexiprecision generalizes $\rlone$ evaluation and best-case performance across two types of uncertainty.  
\subsection{Quantifying Preferences}
Although lexiprecision provides a ordering over a pair of rankings, it does not quantify the magnitude of the preference (i.e. the value of $\preferencefn(\rlx,\rly)$).  
Defining a magnitude allows us to measure the degree of preference, which can then be averaged over multiple requests.  

We can define the magnitude directly as the value of $\drr_i$ and, therefore, defining $\preferencefn(\rlx,\rly)$ as,
\begin{align}
    \rrlexiprecision(\rlx,\rly)&=\drr_{i^*}(\rlx,\rly)
\end{align}
where $i^*$ is defined in Section \ref{sec:algorithms:best-case-lexiprecision}.  This has the advantage of, when $i^*=1$, reproducing the difference in reciprocal rank.  
Under this definition, the magnitude of preferences for higher recall levels will tend to be smaller due to the aggressive discounting in reciprocal rank.  

Alternatively, we can be more conservative in our quantification and just return a constant value based on the preference, defining $\preferencefn(\rlx,\rly)$ as,
\begin{align}
    \sgnlexiprecision(\rlx,\rly)&=\sgn(\drr_{i^*}(\rlx,\rly))
\end{align}
where $i^*$ is defined as above.  
Although the direction of the preference agrees with $\rrlexiprecision$, we discard its magnitude and, as a result, differences at lower ranks are equal to those at higher ranks.  Prior work found that looking at unweighted preference information alone can help with preference sensitivity \cite{diaz:rpp}.

\subsection{Lexicographic Precision as Modeling $\drr_1$}
\label{sec:algorithms:modelingrr}
A different way to interpret lexiprecision is as a method to estimate a high-precision preference between rankings.  Assume that we have some latent preference between two rankings, $\latentpreferencefn(\rlx,\rly)$, that we know to be `high-precision'.  That is, users prefer finding \textit{some} relevant items quickly than \textit{all} relevant items quickly.

One way to model this preference is to inspect the positions of relevant items in $\rlx$ and $\rly$.  From the perspective of `very high precision', observing $\drr_1(\rlx,\rly)>0$ provides significant evidence that $\latentpreferencefn(\rlx,\rly)>0$.  What if we do not observe a preference at the first recall level?  Inspired by Katz's back-off model \cite{katz:backoff}, we inspect the second recall level for evidence of the value of $\latentpreferencefn(\rlx,\rly)$.  If we do not observe a preference, we can progressively back off to higher and higher recall levels.

\begin{figure}
    \includegraphics[width=0.75\linewidth]{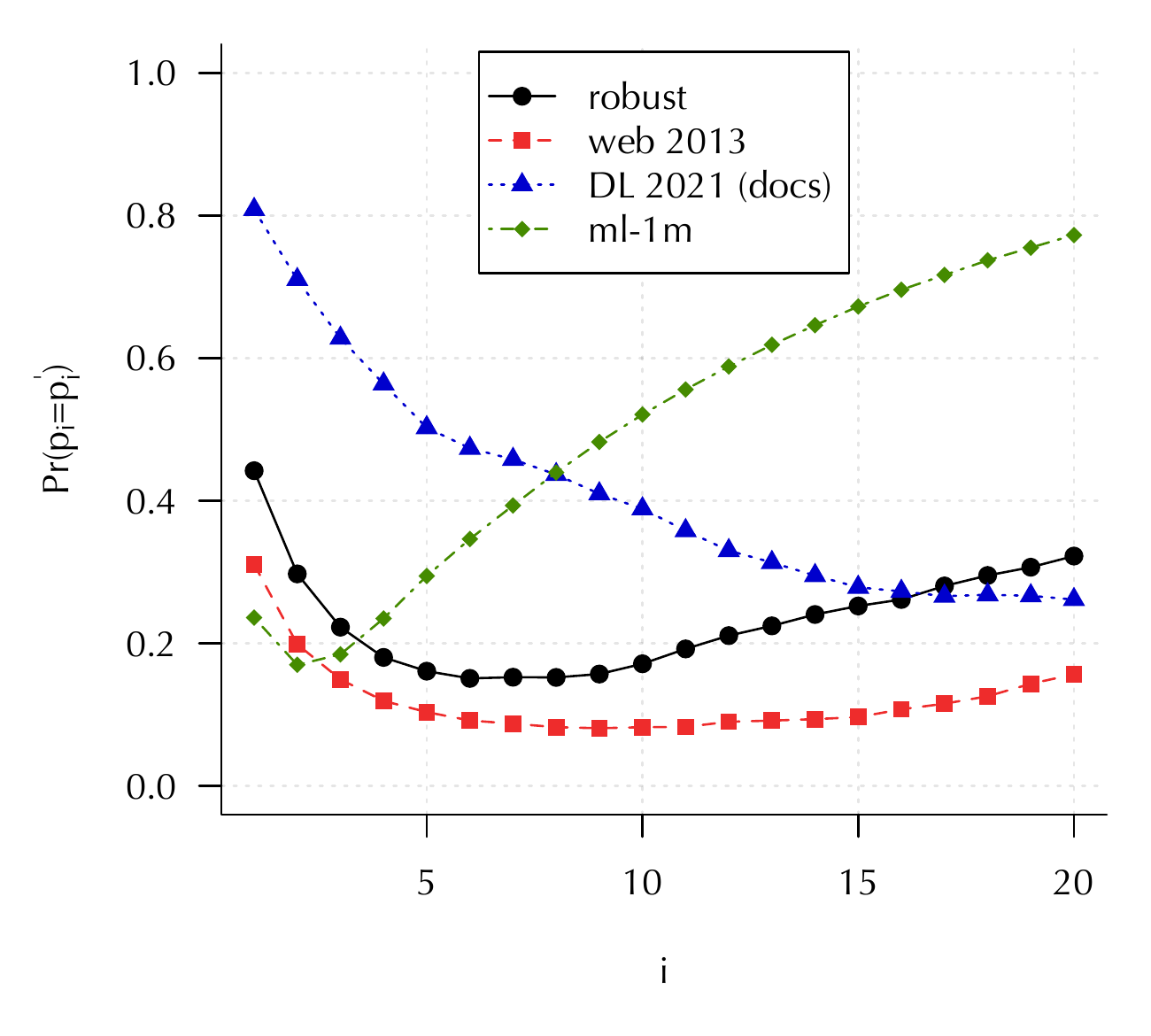}
    \caption{Empirical probability of a tie in position by recall level.  Note that, while Figure \ref{fig:motivation:ptie} measures the probability of a tie for different positions of the highest ranked relevant item (i.e. $\RPx_1$), this figure measures the probability of a tie for different recall levels. }\label{fig:motivation:numties}
\end{figure}

While Section \ref{sec:motivation} demonstrated that $\drr_1(\rlx,\rly)=0$ with high probability, backing off our estimates  works best if, for $i>1$, we expect $\drr_i(\rlx,\rly)=0$ with lower probability.  Using the runs associated with several public benchmarks, we computed $\drr_i$ for all pairs of rankings generated by multiple systems for the same query.  We show the probability of a tie for the first twenty recall levels in Figure \ref{fig:motivation:numties}.  We can see that the number of ties at $\drr_1$ are high, ranging from roughly 20\

Inspecting the number of relevant items retrieved confirms this.  The DL 2021 submissions had $38.74\pm 21.75$ relevant items in their retrievals, compared to web with $53.14\pm47.06$.  Meanwhile, robust submissions had $40.51\pm 41.49$ relevant items retrieval, suggesting much higher variance and ml-1m with $7.46\pm8.57$ relevant items retrieved and much higher variance, leading to more more ties at higher recall levels. 

\begin{figure}
    \includegraphics[width=0.75\linewidth]{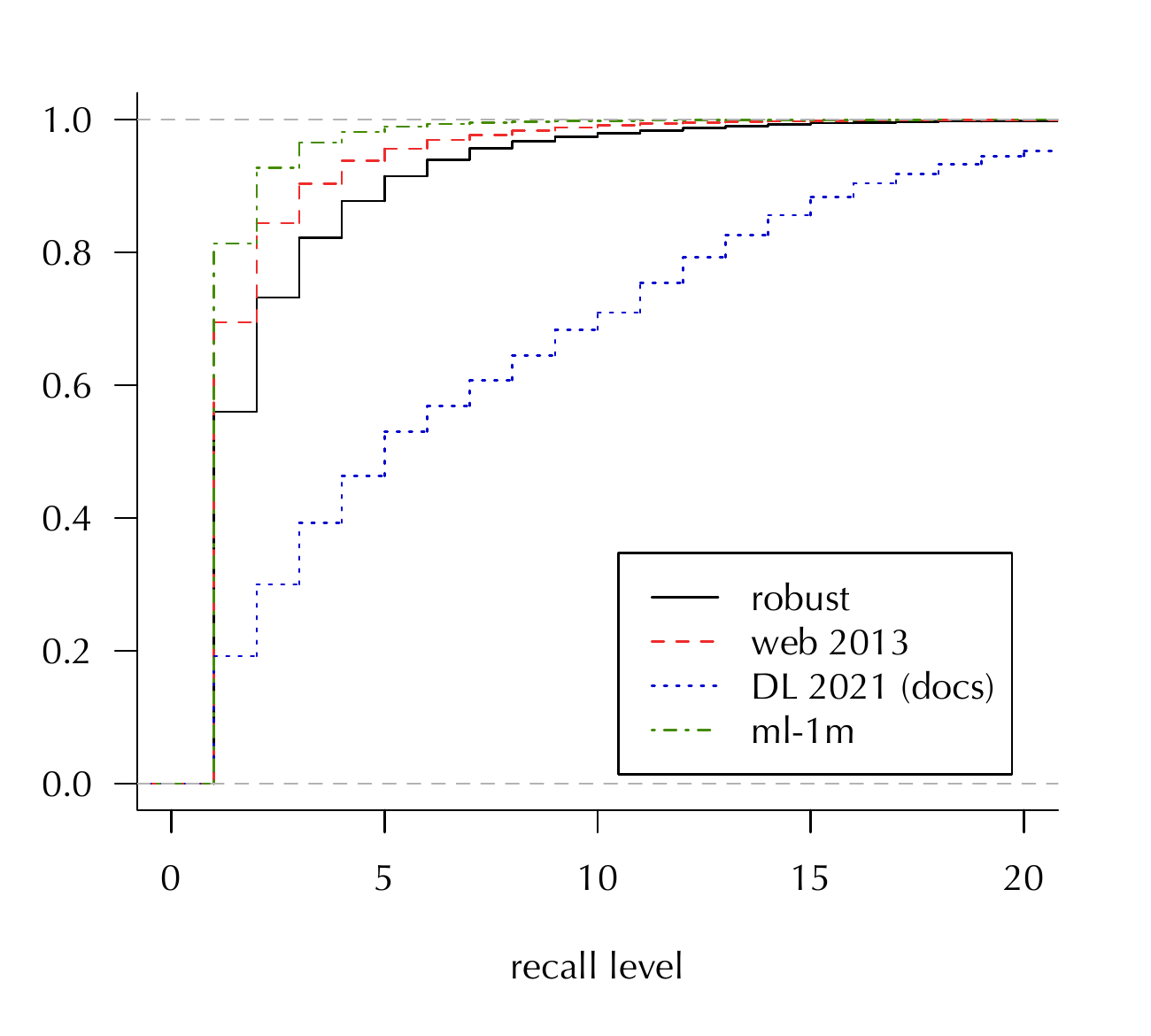}
    \caption{Empirical cumulative distribution function of recall level needed to distinguish systems.}\label{fig:recall-level-depth}
\end{figure}

Given that different benchmarks observed different behaviors for ties amongst recall levels, we need to understand how many recall levels we need to visit before finding evidence for $\latentpreferencefn$.  If a benchmark needs many recall levels but observes many ties at high recall levels, then our model of $\latentpreferencefn$ may be less reliable.  We computed the number of recall levels needed, $i^*$, for each benchmark and plotted the empirical cumulative distribution function in Figure \ref{fig:recall-level-depth}. We find that we need fewer than ten recall levels to capture 90\

\begin{figure}
    \centering

    \begin{subfigure}[b]{0.9\columnwidth}
        \centering
        \includegraphics[width=0.85\linewidth]{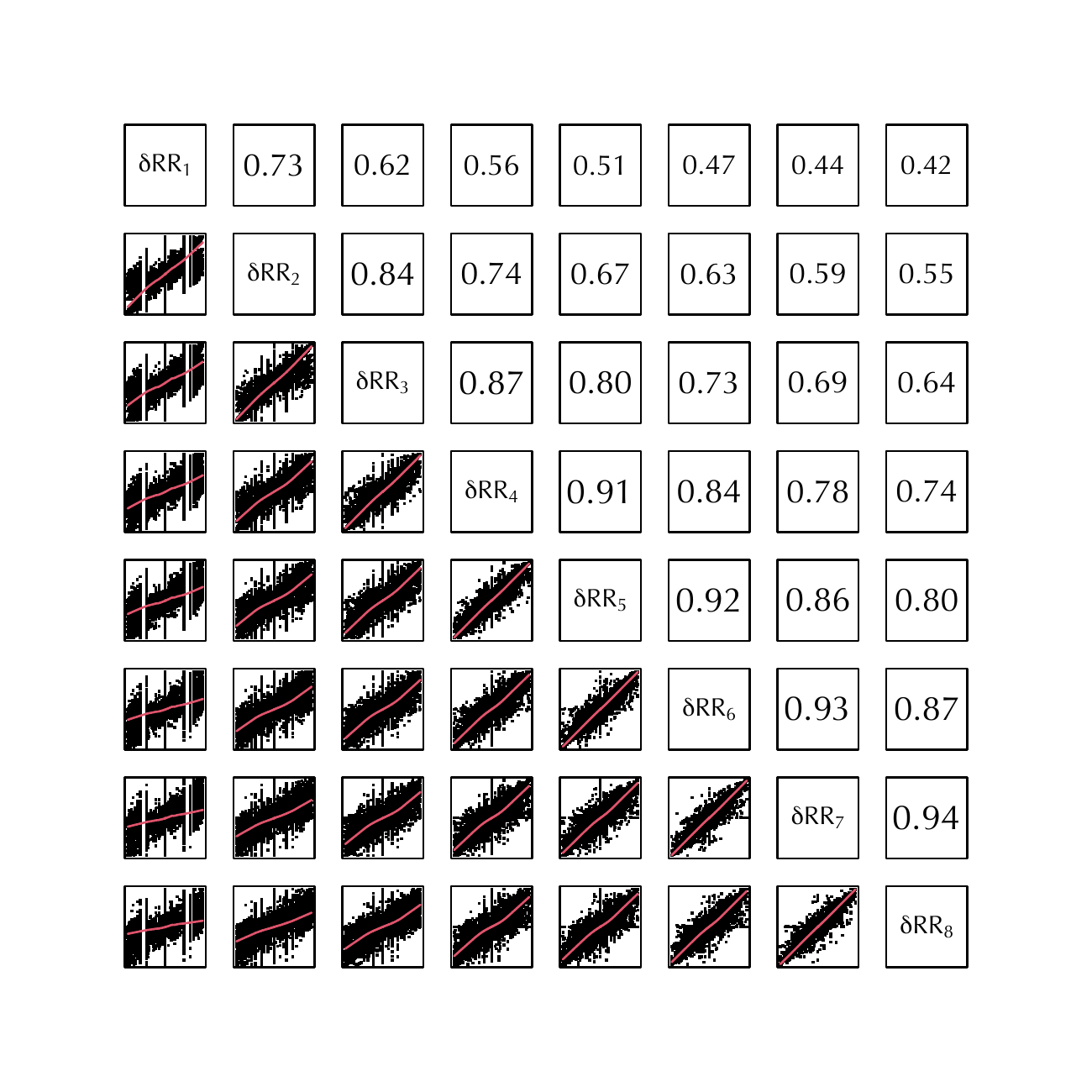}
        \caption{Correlation between $\drr_i$}\label{fig:predi:correlation}
    \end{subfigure}
    
    \begin{subfigure}[b]{0.9\columnwidth}
        \centering
        \includegraphics[width=0.8\linewidth]{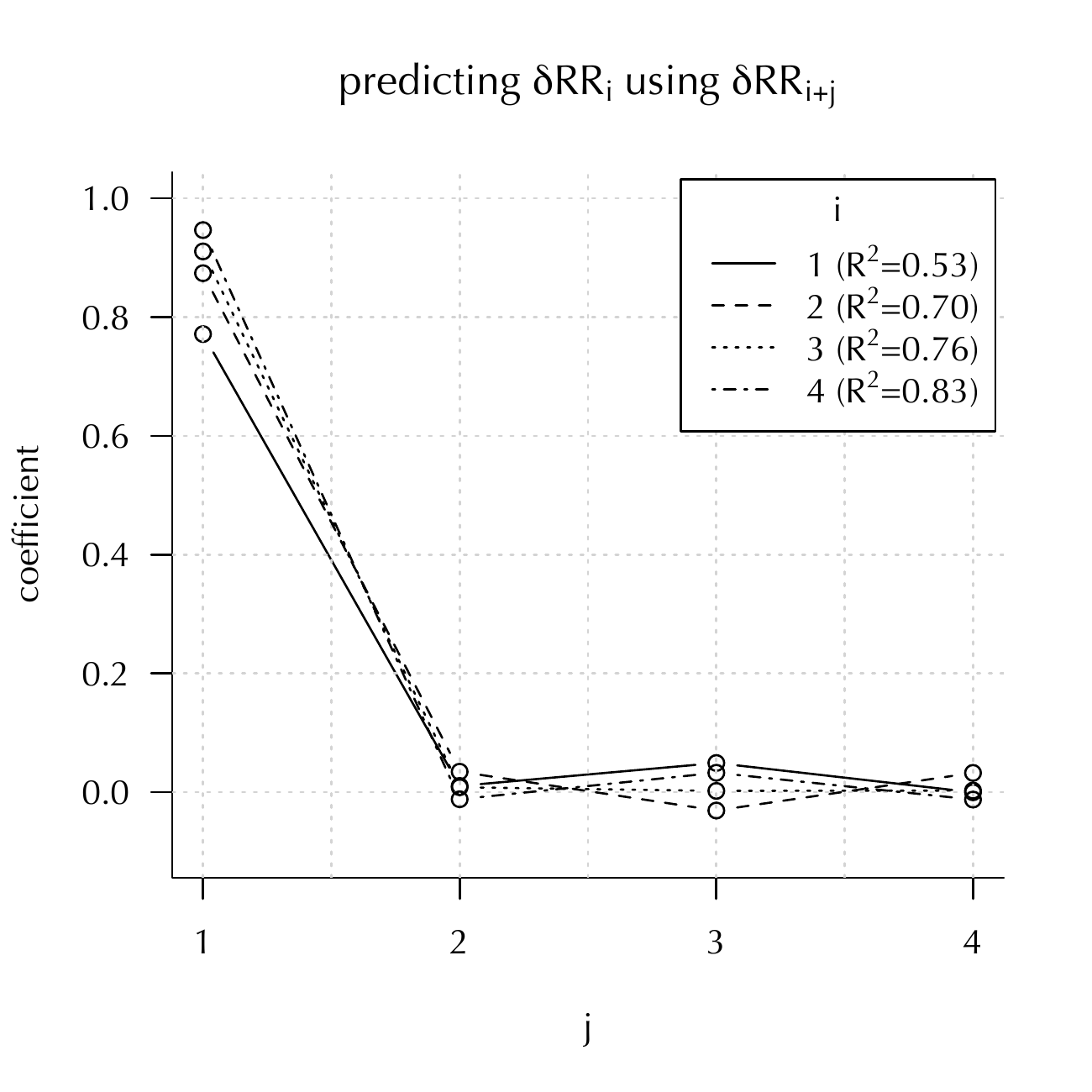}
        \caption{Regression of $\drr_i$ using $\drr_{i+1},\ldots,\drr_{i+4}$ }\label{fig:predi:regression}
    \end{subfigure}
    \caption{Locality of $\drr_i$.  Relationship between the difference in reciprocal rank across recall levels using Robust 2004 runs. (\subref{fig:predi:correlation}) Pearson and linear fit between all pairs of $\drr_i$.  (\subref{fig:predi:regression}) Linear regression of $\drr_i$ using $\drr_{i+1:i+4}$ as independent variables.  Regression shown for $i=\{1,2,3,4\}$.}\label{fig:predi}
\end{figure}

Although our preceding analysis demonstrates that a backoff model of $\latentpreferencefn$ based on lexiprecision will terminate at a reasonable depth, we still need to show that there is locality amongst $\drr_i$.  This means that we ask, if we observe $\drr_1(\rlx,\rly)>0$, how likely is it that $\drr_2(\rlx,\rly)>0$?  $\drr_3(\rlx,\rly)>0$?  If there is high locality amongst $\drr_i$, then information from $\drr_{i+1}$ can help in predicting the true value of $\drr_i$ when it is missing or tied.  Note that, if we observe $\drr_i> 0$ and $\numdocs$ is large, there is absolutely no guarantee that $\drr_{i+1}>0$ since the next ranked relevant items could, in theory, occur anywhere in the range $[\RPx_i+1,\numdocs]$ and $[\RPy_i+1,\numdocs]$.  That said, given the number of ties at recall level 1, we are interested in understanding whether information at other rank positions can provide a way to distinguish tied rankings.  In Figure \ref{fig:predi:correlation}, we computed the Pearson correlation amongst all pairs of $\drr_i$ for $i\in[1,8]$ for the Robust 2004 benchmark.  The fact that correlation between $\drr_i$ and $\drr_{i+j}$ degrades as $j$ increases from 1 demonstrates  that there is indeed high locality.  The implication justifies the use of backoff modeling of $\latentpreferencefn$.

To test this hypothesis explicitly, we fit a linear model of $\drr_1$ using $\drr_2,\ldots, \drr_4$ as independent variables.  We plot the coefficients of the linear regression in the solid line in Figure \ref{fig:predi:regression}.  The substantially larger coefficient on $\drr_2$ indicates that the majority of the predictive power can be found at recall level 2 ($j=1$).  Higher recall levels ($j>1$) are associated with much smaller coefficients.  The actual contributions of higher recall levels are much smaller than this suggests since, because we are operating with reciprocals,  the magnitude of $\drr_i$ shrinks as $i$ grows.  While the colinearity in Figure \ref{fig:predi:correlation} might explain some of this disparity in weights, the locality of individual Pearson correlations and high predictive accuracy means, from a modeling perspective, that a backoff model is justified.    We repeated this analysis for predicting $\drr_2$ from $\drr_3,\ldots,\drr_6$ and similarly for $\drr_3$ and $\drr_4$. 

Similar to our observation when modeling $\drr_1$, these results suggest that the next higher recall level is the most valuable predictor when modeling $\drr_i$ for any specific recall level.  

We repeated this regression analysis for explicitly cascaded data (i.e. only modeling cases when there is a tie at positions $i'<i$) as well as for regressing against the sign of the preference and observed identical findings.  Although we omit those plots due to space constraints, they further support a backoff model intrepretation of lexiprecision.

\section{Methods}
\label{sec:methods}
In previous sections, we theoretically and conceptually connected $\rlone$  to the notion of best-case retrieval evaluation, with a few illustrative empirical results.  In order to rigorously test the viability of lexiprecision, we conducted a series of empirical analyses based on publicly available benchmarking data.\footnote{Code for computing lexiprecision can be found at \url{https://github.com/diazf/pref_eval}.}  
\subsection{Data}
\label{sec:data}
We analyzed the performance of lexiprecision across a variety of retrieval and recommendation tasks.  Specifically, we collected runs submitted to TREC news (Robust 2004, Core 2017 and 2018), web (Web 2009-2014), and deep learning (Deep Learning 2019-2021) tracks as well as several public recommendation tasks \cite{valcarce:recsys-ranking-metrics-journal}.  We present details of these datasets in Table \ref{tab:data}.  

\begin{table}[t]
    \caption{Datasets used in empirical analysis.}\label{tab:data}
{\small
    \begin{tabular}{lcccc}
        \hline
        &   requests    &   runs    &   rel/request &   docs/request    \\
        \hline
        \textbf{news}\\
        robust (2004) &       249 &  110 &    69.93  &    913.82     \\
        core (2017)     &       50  &  75   &   180.04 &        8853.11     \\
        core (2018)     &       50  &  72   &   78.96  &        7102.61     \\
        \textbf{web}\\
        web (2009)      &       50  &  48 & 129.98  &   925.31     \\
        web (2010)      &       48  &  32 & 187.63  &   7013.21     \\
        web (2011)      &       50  &  61 & 167.56  &   8325.07     \\
        web (2012)      &       50  &  48 & 187.36  &   6719.53     \\
        web (2013)      &       50  &  61 & 182.42  &   7174.38     \\
        web (2014)      &       50  &  30 & 212.58  &   6313.98     \\
        \textbf{deep}\\
        deep-docs (2019)        &       43  &  38 & 153.42 & 623.77     \\
        deep-docs (2020)        &       45  &  64 & 39.27  & 99.55     \\
        deep-docs (2021)        &       57  &  66 & 189.63  & 98.83     \\
        deep-pass (2019)        &       43  &  37 & 95.40  & 892.51     \\
        deep-pass (2020)        &       54  &  59 & 66.78  & 978.01     \\
        deep-pass (2021)        &       53  &  63 & 191.96  & 99.95     \\
        \textbf{recsys}\\
        movielens   &       6005    &  21 & 18.87  & 100.00\\
        libraryThing &   7227    &   21  &   13.15   &  100.00\\
        beerAdvocate &   17564    &   21  &   13.66   & 99.39\\
\hline
\end{tabular}
}
\end{table}

\subsection{Analyses}
Our empirical analyses were founded on two core questions,
\begin{inlinelist}
    \item how empirically correlated are lexiprecision and $\rlone$ metrics, and
    \item how much more robust is lexiprecision than $\rlone$ metrics.
\end{inlinelist}
Because of its widespread adoption in the research community, we will use reciprocal rank for analyses.
In order to answer the first question, we conducted experiments designed to predict the agreement between lexiprecision and $\rlone$ metrics under different conditions.  We considered two types of agreement.  \textit{Agreement in ranking preference} tests whether $\rlx\lexiprecisionpref\rly$ agrees with $\rlx\rrpref\rly$.  Because lexiprecision is substantially more sensitive than $\rlone$ metrics, we only consider situations where $\drr_1(\rlx,\rly)\neq0$.  Because $\sgnlexiprecision$ and $\rrlexiprecision$ always agree in sign, we will only show results for one of the metrics when computing ranking agreement.  \textit{Agreement in system preference} tests whether $\expectation{\query\sim\queries}{\preferencefn_{\lexiprecision}(\rlx_\query,\rly_\query)}$ agrees in sign with $\expectation{\query\sim\queries}{\preferencefn_{\rr}(\rlx_\query,\rly_\query)}$.  This measures whether our choice of $\rrlexiprecision$ or $\sgnlexiprecision$ affects its correlation with reciprocal rank.  Agreement is measured as a percentage of preferences agreed upon. 

In order to assess the robustness of lexiprecision, we measure the number of ties observed amongst pairs of rankings  and discriminative power.  We claim that a robust approach has fewer ties and higher discriminative power.   For discriminative power, we adopt Sakai's approach of measuring the number of statistically significant differences between runs \cite{sakai:metrics}, using both Tukey's honestly significant difference (HSD) test    \cite{carterette:multiple-testing} and classic paired test to compute $p$-values.  The paired test uses the Student's $t$-test for reciprocal rank and $\rrlexiprecision$ \cite{smucker:tests}; and the binomial test for $\sgnlexiprecision$.

\section{Results}
\label{sec:results}

\begin{table}
    \caption{Ranking agreement between $\drr_1$ and preferences based on the positions of the last $\numrel-1$ relevant items.  The computation of $\sgnlexiprecision$ in the table is based on the $\numrel-1$ positions of relevant items after the top-ranked relevant item. }\label{tab:rr1prediction}
    {\small
    \begin{tabular}{lc|c}
        \hline
        & $\sgnlexiprecision$ & $\drr_2$ \\\hline
        \textbf{news} & & \\
        robust (2004)	&	\textbf{85.78}	&	83.44	\\
        core (2017)	&	\textbf{89.23}	&	87.30	\\
        core (2018)	&	\textbf{88.01}	&	86.58	\\
        \textbf{web} & & \\
        web (2009)	&	\textbf{85.87}	&	84.79	\\
        web (2010)	&	\textbf{87.29}	&	85.41	\\
        web (2011)	&	\textbf{88.91}	&	87.54	\\
        web (2012)	&	\textbf{87.22}	&	85.45	\\
        web (2013)	&	\textbf{86.51}	&	84.45	\\
        web (2014)	&	\textbf{88.02}	&	85.82	\\
        \textbf{deep} & & \\
        deep-docs (2019)	&	\textbf{86.56}	&	83.10	\\
        deep-docs (2020)	&	\textbf{83.73}	&	79.34	\\
        deep-docs (2021)	&	\textbf{92.41}	&	89.78	\\
        deep-pass (2019)	&	\textbf{90.45}	&	88.87	\\
        deep-pass (2020)	&	\textbf{92.86}	&	91.08	\\
        deep-pass (2021)	&	\textbf{91.97}	&	90.14	\\
        \textbf{recsys} & & \\
        ml-1M (2018)	&	\textbf{78.90}	&	77.56	\\
        libraryThing (2018)	&	\textbf{66.50}	&	66.08	\\
        beerAdvocate (2018)	&	\textbf{58.84}	&	58.25	\\
        \hline
    \end{tabular}
    }
\end{table}

\begin{figure}[t]
    \centering
    \begin{subfigure}[b]{0.9\columnwidth}
    \centering
        \includegraphics[width=1.55in]{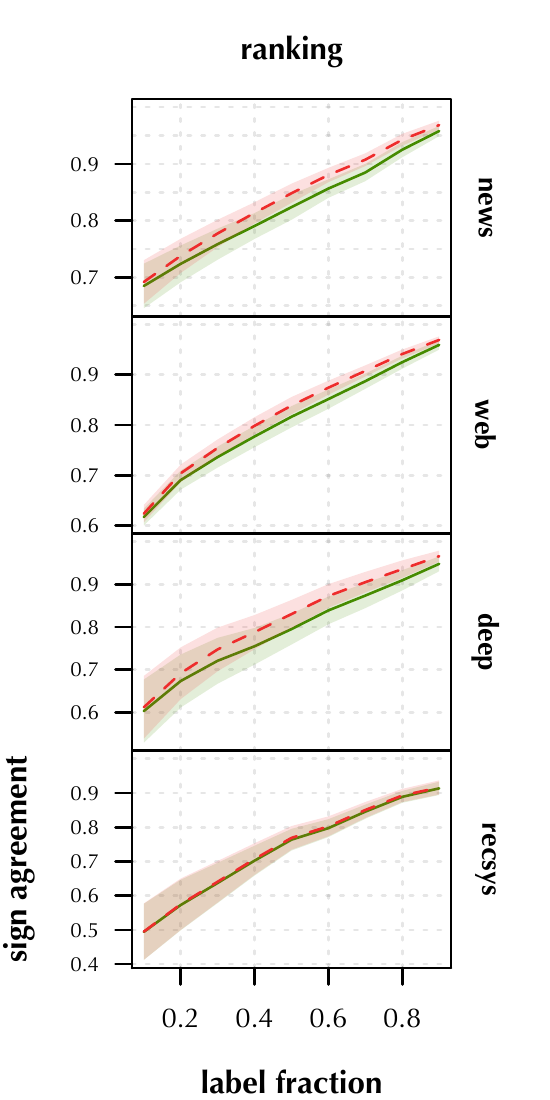}\includegraphics[width=1.55in]{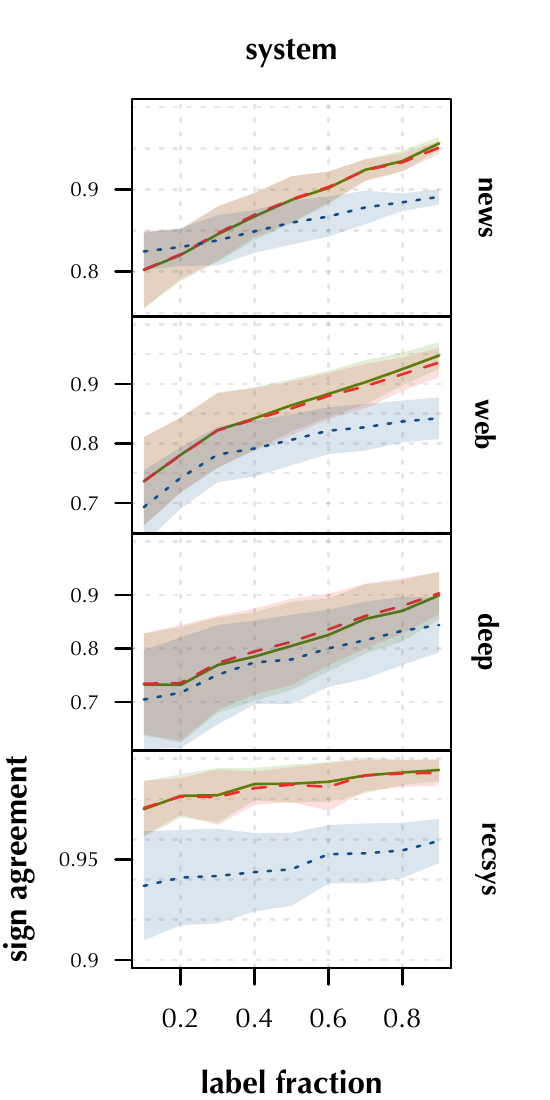}
        \caption{Removing labels. }\label{fig:degradation:rr:labels}
    \end{subfigure}

    \begin{subfigure}[b]{0.9\columnwidth}
        \centering
    
    \includegraphics[width=1.55in]{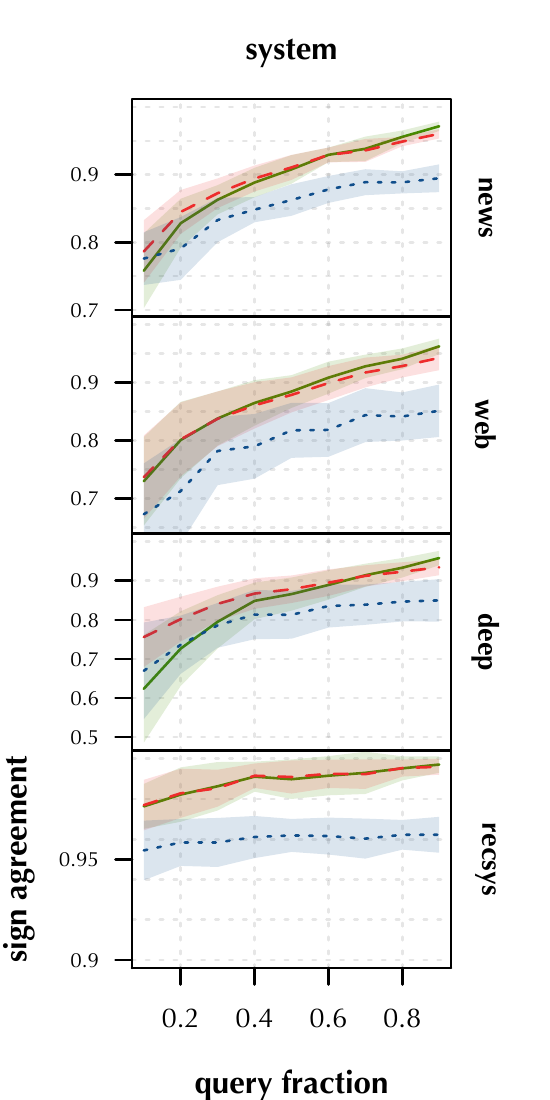}
    
        \caption{Removing queries. }\label{fig:degradation:rr:queries}
    \end{subfigure}
    \caption{Preference agreement with $\drr_1$ with full data. Labels and requests removed randomly.  Results averaged across ten samples. Solid green lines: $\drr_1$ with incomplete information.  Dashed red lines: $\rrlexiprecision$ with incomplete information.  Dotted blue lines: $\sgnlexiprecision$ with incomplete information.  Shaded areas: one standard deviation across samples.  Ranking agreement with incomplete labels for $\sgnlexiprecision$ is identical to $\rrlexiprecision$ and omitted for clarity.}
\end{figure}

\subsection{Correlation with Reciprocal Rank}
By construction, we know that $\drr_1(\rlx,\rly)>0\implies\lexiprecision(\rlx,\rly)>0$ and, so, the correlation between the two will be  high.  We can further test this by comparing how well lexiprecision predicts a ground truth preference between rankings based on $\drr_1$.  

In our first analysis, given an observed $\drr_1(\rlx,\rly)\neq0$, we measure the ability of lexiprecision and reciprocal \textit{based only on $\numrel-1$ subsequent recall levels} to predict the sign of $\drr_1(\rlx,\rly)$.  That is, we use $\drr_1(\rlx,\rly)$ as a target value and compute  $\drr_1$ and $\lexiprecision$ using suffixes $\RPx_{2:\numrel}$ and $\RPy_{2:\numrel}$. Although artificial, this analysis provides an indication of the predictive value gained through cascaded modeling (as opposed to just looking at the top-ranked relevant item).  We present the results in Table \ref{tab:rr1prediction}.  As we can see, lexiprecision consistently agrees more with the target (masked) $\drr_1$ than $\drr_1$ of the suffix across all datasets, indicating that the additional information in higher recall levels can be used to predict the target (masked) $\drr_1$.  This agrees with our preliminary analysis in Section \ref{sec:algorithms:modelingrr}.

We can also test the relationship between reciprocal rank and lexiprecision by measuring the agreement under incomplete information.  Specifically, we consider removing either labels (treating unlabeled items as non-relevant) or requests (i.e. queries or users).  We then measure the agreement between preferences with incomplete data and $\drr_1$ on complete data (i.e. all requests and labels).  Methods that agree more with reciprocal rank on complete data  are considered more correlated.  We present results for ranking and system agreement when removing labels (Figure \ref{fig:degradation:rr:labels}) and queries (Figure \ref{fig:degradation:rr:queries}).  Across all conditions, we observe that the $\rrlexiprecision$ has as high or slightly higher agreement with $\drr_1$ with complete information than $\drr_1$ with incomplete information.  This means that $\rrlexiprecision$ can accurately predict  $\drr_1$ with complete information as well or better than using reciprocal rank.  Moreover, we observed that $\sgnlexiprecision$ shows weaker system agreement which occurs because its magnitude does not decay with rank position and, therefore, resulting averages are inconsistent with averages of position-discounted reciprocal rank values.

\begin{table}
    \caption{Percentage of ties between pairs of rankings from two systems for the same request.  We collapse $\rrlexiprecision$ and $\sgnlexiprecision$ for clarity. }\label{tab:numties}
    {\small    
    \begin{tabular}{lc|c}
        \hline
        
        & $\rrlexiprecision$, $\sgnlexiprecision$ & $\drr_1$ \\\hline
        \textbf{news} & & \\
        robust (2004)	&	\textbf{0.39}	&	44.22	\\
        core (2017)	&	\textbf{0.23}	&	48.50	\\
        core (2018)	&	\textbf{1.72}	&	31.43	\\
        \textbf{web} & & \\
        web (2009)	&	\textbf{4.93}	&	15.13	\\
        web (2010)	&	\textbf{0.61}	&	25.85	\\
        web (2011)	&	\textbf{1.02}	&	41.99	\\
        web (2012)	&	\textbf{0.34}	&	34.01	\\
        web (2013)	&	\textbf{0.83}	&	31.09	\\
        web (2014)	&	\textbf{0.64}	&	41.93	\\
        \textbf{deep} & & \\
        deep-docs (2019)	&	\textbf{1.06}	&	68.45	\\
        deep-docs (2020)	&	\textbf{2.43}	&	73.99	\\
        deep-docs (2021)	&	\textbf{0.23}	&	80.84	\\
        deep-pass (2019)	&	\textbf{2.63}	&	56.89	\\
        deep-pass (2020)	&	\textbf{2.58}	&	50.30	\\
        deep-pass (2021)	&	\textbf{1.32}	&	47.41	\\
        \textbf{recsys} & & \\
        ml-1M (2018)	&	\textbf{3.38}	&	21.39	\\
        libraryThing (2018)	&	\textbf{16.48}	&	25.85	\\
        beerAdvocate (2018)	&	\textbf{41.73}	&	45.72	\\
        \hline
    \end{tabular}
    }
\end{table}

\begin{table*}[t]
    \centering
    \caption{Percentage of run differences detected at $p<0.05$. \better{Red}: better than reciprocal rank. \textbf{Bold}: best for an evaluation setting. }\label{tab:sensitivity}
    {\small
        \begin{subtable}[b]{0.30\linewidth}
        \caption{Tukey's HSD test}\label{tab:sensivity:hsd}
        \begin{tabular}{lcc|c}
            \hline
            &   rrLP  & sgnLP & RR  \\
            \hline
            \textbf{news}&&&\\
            robust (2004)	&	\best{27.42}	&	\better{27.34}	&	23.55	\\
            core (2017)	&	\best{17.41}	&	14.67	&	15.03	\\
            core (2018)	&	\better{28.60}	&	\best{31.42}	&	27.39	\\
            \textbf{web}&&&\\
            web (2009)	&	23.85	&	\best{28.28}	&	24.11	\\
            web (2010)	&	\best{18.95}	&	13.51	&	18.35	\\
            web (2011)	&	\best{14.70}	&	10.22	&	13.83	\\
            web (2012)	&	\best{13.39}	&	11.61	&	13.21	\\
            web (2013)	&	5.85	&	5.79	&	\textbf{6.07}	\\
            web (2014)	&	\best{20.00}	&	11.72	&	18.85	\\
            \textbf{deep}&&&\\
            deep-docs (2019)	&	\better{8.25}	&	\best{19.20}	&	6.97	\\
            deep-docs (2020)	&	\best{5.26}	&	\better{3.47}	&	2.88	\\
            deep-docs (2021)	&	\better{6.39}	&	\best{11.19}	&	4.48	\\
            deep-pass (2019)	&	\better{16.52}	&	\best{18.47}	&	13.21	\\
            deep-pass (2020)	&	\better{37.46}	&	\best{40.91}	&	28.35	\\
            deep-pass (2021)	&	\better{24.07}	&	\best{24.78}	&	20.38	\\
            \textbf{recsys}&&&\\
            ml-1M (2018)	&	\better{81.43}	&	\best{90.95}	&	80.00	\\
            libraryThing (2018)	&	93.81	&	\best{96.67}	&	93.81	\\
            beerAdvocate (2018)	&	\better{92.38}	&	\best{96.19}	&	90.95	\\
            \hline
            \end{tabular}
        \end{subtable}
    }\hspace{1in}{\small
\begin{subtable}[b]{0.30\linewidth}
    \caption{Paired test with Bonferroni correction }\label{tab:sensivity:tt}
    \begin{tabular}{lcc|c}
        \hline
            &   rrLP  & sgnLP & RR  \\
            \hline
            \textbf{news}&&&\\
            robust (2004)	&	\better{26.22}	&	\best{27.36}	&	21.45	\\
            core (2017)	&	\best{16.22}	&	11.35	&	11.53	\\
            core (2018)	&	\better{29.30}	&	\best{31.73}	&	27.03	\\
            \textbf{web}&&&\\
            web (2009)	&	\better{23.76}	&	\best{25.18}	&	23.49	\\
            web (2010)	&	\best{18.55}	&	9.27	&	17.74	\\
            web (2011)	&	\best{12.30}	&	6.94	&	9.73	\\
            web (2012)	&	\best{11.97}	&	10.11	&	11.35	\\
            web (2013)	&	\best{4.75}	&	\better{4.64}	&	4.32	\\
            web (2014)	&	\best{15.86}	&	7.13	&	14.02	\\
            \textbf{deep}&&&\\
            deep-docs (2019)	&	\better{11.66}	&	\best{16.36}	&	5.69	\\
            deep-docs (2020)	&	\best{2.33}	&	\better{1.79}	&	0.60	\\
            deep-docs (2021)	&	\better{3.73}	&	\best{9.14}	&	3.03	\\
            deep-pass (2019)	&	\better{15.02}	&	\best{17.42}	&	10.36	\\
            deep-pass (2020)	&	\better{39.04}	&	\best{39.45}	&	28.00	\\
            deep-pass (2021)	&	\best{23.55}	&	\better{20.99}	&	16.79	\\
            \textbf{recsys}&&&\\
            ml-1M (2018)	&	90.00	&	\best{92.38}	&	90.48	\\
            libraryThing (2018)	&	\better{97.14}	&	\best{97.62}	&	96.67	\\
            beerAdvocate (2018)	&	94.76	&	\best{96.67}	&	94.76	\\
            \hline
            \end{tabular}
        \end{subtable}
}
\end{table*}

\subsection{Sensitivity}
In Section \ref{sec:motivation}, we motivated our work by showing that $\rlone$ metrics theoretically and empirically suffer from ceiling effects.  The primary instrument we used to determine this was the probability of ties between rankings.  In Table  \ref{tab:numties}, we present the percentage of tied rankings from different systems for the same request.  As predicted by our analysis in Section \ref{sec:algorithms:numties}, lexiprecision has substantially fewer ties because this only happens when two rankings place relevant items in exactly the same positions.

\begin{figure}
    \centering
    \includegraphics[width=1.75in]{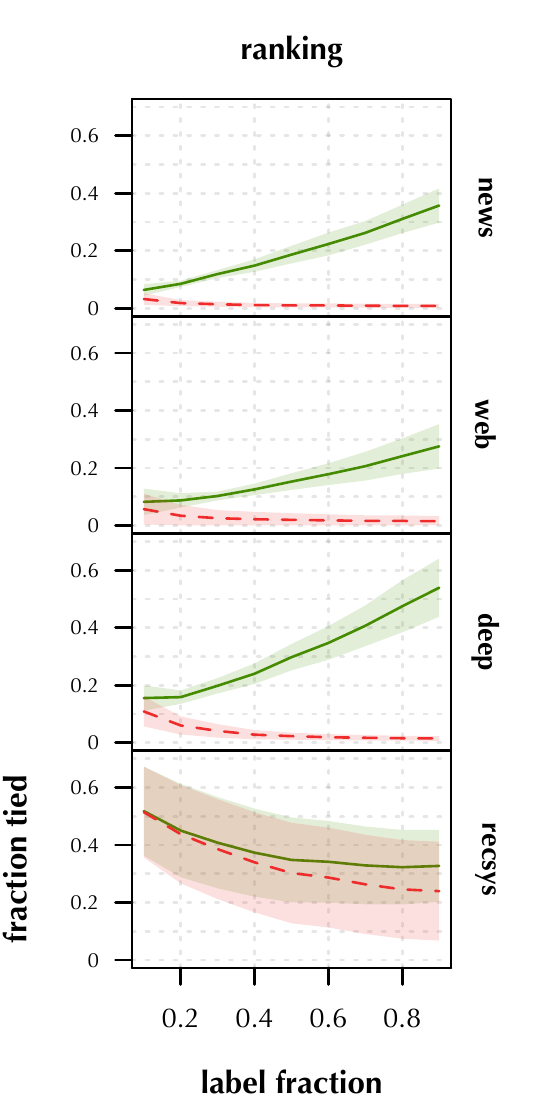}
    \caption{Number of ties as labels are removed randomly.  Results are averaged across ten samples. Solid green lines: $\drr_1$ with incomplete information.  Dashed red lines: $\rrlexiprecision$ with incomplete information.  Shaded areas: one standard deviation across samples.  Number of ties with incomplete labels for $\sgnlexiprecision$ is identical to $\rrlexiprecision$ and omitted for clarity.}\label{fig:numties}
\end{figure}

In Section \ref{sec:algorithms:numties}, we showed that lexiprecision implicitly and exponentially increased its fidelity as the number of relevant items $\numrel$ increased, while $\rlone$ would quickly suffer from ties.  In Figure \ref{fig:numties}, we show the number of tied rankings as a function of incomplete labels.  This allows us to see trends with respect to $\numrel$.  Across our three retrieval benchmark sets, we see the growth in number of ties for $\rlone$ as $\numrel$ increases; meanwhile,  they shrink for lexiprecision.  The drop in ties for recommender systems benchmarks suggests that, as described in Section \ref{sec:algorithms:modelingrr}, rankings contain very few relevant items and, as a result, removing labels will result in no relevant items present and increasingly tied rankings.

While the number of ties indicates that $\rlone$ might not be able to distinguish systems, for a large enough sample of requests, a metric might still be good enough to distinguish systems.  A different approach to measuring the discriminative power of an evaluation method is to count the number of differences that are statistically significant \cite{sakai:metrics}.  When we compare the percentage of pairs registering a statistically significant difference (Table \ref{tab:sensitivity}), both $\rrlexiprecision$ and $\sgnlexiprecision$ outperform reciprocal rank, often by a very large margin.  This indicates that the number of ties indeed hurts the ability of reciprocal rank to detect significant differences, while both variants of lexiprecision are much more sensitive.

\section{Discussion}
\label{sec:discussion}
Our results demonstrate that our  lexiprecision variants capture the properties of $\rlone$ while substantially increasing the ability to distinguish systems under the same best-case evaluation assumptions.  

Practitioners and evaluators need to assess  whether the assumptions behind $\rlone$ metrics, including reciprocal rank, or lexiprecision or any other evaluation scheme are aligned with the use case.  If a retrieval environment supports the assumptions behind $\rlone$ metrics, \textit{including ties}, then, by all means, they should be used to assess performance.  However, in Section \ref{sec:algorithms:best-case}, we raised several reasons why uncertainty over recall requirements and psychological relevance suggest that $\rlone$ metrics make quite strong assumptions not realized in most retrieval settings.  We  designed lexiprecision to operate as conservatively as possible, preserving any preference from $\rlone$ metrics and only acting to break ties. 

Although $\rlone$ metrics and lexiprecision agree perfectly when there is only one relevant item, this does not mean that all situations where we have a single judged relevant item should adopt a metric like reciprocal rank.  For example, the MSMARCO dataset \cite{msmarco} includes requests and very sparse labels; the majority of requests have one \textit{judged} relevant  item.   One might be tempted to use reciprocal rank but \citet{arabzadeh:shallow-pooling-for-sparse-labels} demonstrate that this would obscure the multitude of \textit{unjudged} relevant items (of which there are many).  This hurts efficacy of best-case retrieval evaluation \textit{including reciprocal rank}, as shown in Figures \ref{fig:degradation:rr:labels} and \ref{fig:numties}.  Recommendation tasks have similar issues with sparsity due in part to it being more difficult for a third party to assess the relevance of personalized content and to the difficulty in gathering explicit feedback.  Labels derived from behavioral feedback in general suffer from similar sparsity \cite{bendersky:sparse-fb}.  In this respect, we echo the call from \citet{arabzadeh:shallow-pooling-for-sparse-labels} to make labeling practices across all of these domains much more robust.  Given the observation of \citet{voorhees:too-many-relevants} that better labeling can result in less informative evaluation, we need to also develop more sensitive evaluation schemes such as lexiprecision.

Finally, this study has introduced a new preference-based evaluation method for $\rlone$ metrics.  As such, our focus has been on developing an understanding for comparing pairs of rankings and systems.  We do not claim that lexiprecision itself is a metric and emphasize that we use it for comparing two rankings or systems.  As such, although we address some concerns with reciprocal rank raised by \citet{ferrante:interval-scale-metrics},  we do not make claims about lexiprecision being an interval measure.  That said, the total ordering shown in Figure \ref{fig:hasse} suggests that there may be version of lexiprecision that can indeed be represented as an interval measure.

\section{Conclusion}
\label{sec:conclusion}
Motivated by ceiling effects in evaluation with reciprocal rank, we have attempted to increase our understanding of the metric and designed a well-grounded mitigation to conducting best-case retrieval evaluation.  We have shown that lexiprecision can effectively address the limitations of reciprocal rank in retrieval evaluation. Our results highlight the importance of considering the effects of tie-breaking in the evaluation process and provide a method for conducting more reliable best-case retrieval evaluation.  Given the use of retrieval metrics---including reciprocal rank---outside of information retrieval contexts, we believe these contributions will be relevant to a researchers in the broader research community.   

\bibliographystyle{ACM-Reference-Format}
\bibliography{fdiaz.bib}
\end{document}